\documentclass[useAMS,usenatbib,usegraphicx]{mn2e}
\usepackage{color}
\definecolor{dark-blue}{rgb}{0.15,0.15,0.4}
\usepackage[colorlinks]{hyperref}
\hypersetup{citecolor=dark-blue, linkcolor=dark-blue}
\usepackage{amsmath}
\usepackage{amsfonts}
\usepackage{amssymb}
\usepackage{times}
\usepackage[T1]{fontenc}
\usepackage{aecompl}

\title[X-rays from LBAs]{Enhanced X-ray emission from Lyman Break Analogues and a Possible $L_{\rm X}$--SFR--Metallicity Plane}
\author[M. Brorby, P. Kaaret, A. Prestwich, and F. Mirabel]{M. Brorby$^{1}$\thanks{E-mail:
matthew-brorby@uiowa.edu}, P. Kaaret$^{1}$, A. Prestwich$^{2}$, and I.~F. Mirabel$^{3,4}$\\
$^{1}$Department of Physics and Astronomy, University of Iowa, Iowa City, IA 52242\\
$^{2}$Harvard-Smithsonian Center for Astrophysics, 60 Garden street, Cambridge, MA 02138\\
$^{3}$Instituto de Astronom\'{i}a y F\'{i}sica del Espacio (IAFE), UBA-CONICET, CC 67, Suc. 28, (C1428ZAA), Buenos Aires, Argentina\\
$^{4}$CEA-Saclay, IRFU/DSM/Service d'Astrophysique, 91191 Gif-sur-Yvette, France}
\begin{document}

\pagerange{\pageref{firstpage}--\pageref{lastpage}} \pubyear{2015}

\maketitle


\label{firstpage}

\begin{abstract}
The source of energetic photons that heated and reionized the early Universe remains uncertain. Early galaxies had low metallicity and recent population synthesis calculations suggest that the number and luminosity of high-mass X-ray binaries are enhanced in star-forming galaxies with low metallicity, offering a potentially important and previously overlooked source of heating and reionization. Lyman break analogue (LBA) galaxies are local galaxies that strongly resemble the high-redshift, star-forming Lyman Break Galaxies and have been suggested as local analogues to these metal-deficient galaxies found in the early Universe. We studied a sample of ten LBAs in order to measure the relation between star formation rate and X-ray luminosity. We found that for LBAs with metallicities in the range $12+\log_{10}({\rm O/H}) = 8.15-8.80$, the $L_X-$SFR relation was $\log_{10} (L_X/{\rm SFR}\, {[\rm erg\ s^{-1}\ M_{\sun}^{-1}\ yr]}) = 39.85(\pm 0.10)$ in the $0.5-8$~keV band with a dispersion of $\sigma = 0.25$~dex. This is an enhancement of nearly a factor of $2$ in the $L_{0.5-8\text{keV}}$--SFR relation relative to results for nearby, near-solar metallicity galaxies. The enhancement is significant at the 98.2\% level ($2.4\sigma$). 
Our enhanced $L_X/{\rm SFR}$ relation is consistent with the metallicity-dependent predicted value from population synthesis models. We discuss the possibility of a $L_X$--SFR--Metallicity plane for star-forming galaxies. These results are important to our understanding of reionization and the formation of early galaxies. 
\end{abstract}


\section{Introduction}\label{sect:intro}
In the early Universe $(z>6)$, X-rays from compact sources might have had an important contribution to the heating of the intergalactic medium (IGM) during the Epoch of Reionization~\citep{shull1985,haardt1996,mirabel2011,mcquinn2012,mesinger2013}. The direct study of these high-redshift X-ray sources is untenable due to the spatial resolution and large observation times that would be required. Instead, we study local analogs to the first, low-metallicity galaxies in the early Universe to infer their properties. Blue compact dwarf (BCD) galaxies have been suggest as the best local analogs to such galaxies~\citep{kunth2000}. It has been found that the BCDs have X-ray emission dominated by high mass X-ray binaries (HMXBs)~\citep{thuan2004}. Many recent studies \citep{mapelli2010,kaaret2011,prestwich2013,brorby2014,douna2015} have found enhanced populations of HMXBs in the extremely metal poor galaxies relative to star formation rate (SFR).
These results match predictions from simulations done by Linden~et~al.~(2010) who showed a dramatic increase in bright HMXBs below 20 percent solar metallicity. Linden~et~al.~(2010) explain that the mechanism driving this bright HMXB population increase is due to an increase in the fraction of binaries accreting through Roche lobe overflow (RLO) as opposed to wind accretion systems.

We further explore the relation between X-ray emission and star formation rate for high-redshift, star-forming galaxies. As discussed by \cite{kunth2000}, it is not certain that all early galaxies were small, metal-deficient galaxies, for which the BCDs are analogues. Instead, there may have been larger, gas-rich proto-galaxies starting to form during the Epoch of Reionization, particularly at later times. These objects would have properties similar to the galaxies that have been observed using the Lyman break technique \citep{heckman2005,heckman2011,hoopes2007,overzier2008}. Again, because these galaxies are undetectable in X-rays, except for the longest of observation times, we turn to local analogues. 

In this study, we observed a sample of ten Lyman break analogues (LBAs). These galaxies are part of a rare population of local galaxies which have properties that match the high-redshift, star-forming Lyman break galaxies (LBGs) with regards to stellar mass, star formation rate, and metallicity~\citep{heckman2005}. We measured the X-ray flux for each galaxy in our sample and converted this to a luminosity (see Section~\ref{sect:procedure} for details). Star formation rates were calculated from IR and UV data, as described in Section~\ref{sect:SFR}. \cite{mineo2012}~(hereafter, M12) provided a linear fit to the X-ray luminosity plotted against SFR for their sample of near-solar metallicity galaxies. They found a relation of,
\begin{equation}\label{eqn:mineo1}
L_{0.5-8\text{keV}}(\text{erg s}^{-1}) = 2.61\times 10^{39} \text{ SFR}\ (\text{M}_{\sun} \text{ yr}^{-1}),
\end{equation}
which has been used in simulations to estimate the X-ray contribution to heating and reionization in the early Universe~\citep[e.g.,][]{mesinger2013, power2013}. Equation~(\ref{eqn:mineo1}) is the relation for resolved galaxies. For unresolved galaxies, \cite{mineo2012} estimated that including the diffuse emission increased the X-ray luminosity such that,
\begin{equation}
L_{0.5-8\text{keV}}(\text{erg s}^{-1}) \approx 3.7\times 10^{39} \text{ SFR}\ (\text{M}_{\sun} \text{ yr}^{-1}).
\end{equation}
M12 employ a specific SFR cutoff of ${\rm SFR}/M_\star > 1\times 10^{-10}~{\rm yr}^{-1}$ for their sample selection in order to statistically select for HMXB-dominated starburst galaxies. The sample of LBA galaxies discussed in this paper also meet this criterion (see Table~\ref{tab:results}).

As with the blue compact dwarf galaxies, the LBA X-ray luminosity is expected to be enhanced with respect to star formation rate due to relatively low metal content \citep{fragos2013a}. The sample of galaxies we use have gas-phase metallicities in the range $8.15 < 12+\log_{10}(\rm O/H) < 8.80$. In this range, population synthesis predicts at most a factor of a few enhancement of X-ray emission \citep{fragos2013b} and observational studies of ULX populations have confirmed this prediction \citep[e.g.,][]{mapelli2010,prestwich2013}. Determining the $L_X$/SFR relation for this sample of analogues will allow us to better understand the role of HMXBs in the early Universe.

In Section~\ref{sect:sample}, we outline the selection constraints for our sample of LBAs. The observations used to derive $L_X$, SFR, and $12+\log({\rm O/H})$ are described in Section~\ref{sect:observations}. In Section~\ref{sect:procedure}, we provide detailed descriptions about the methods used to derive X-ray luminosities from the Chandra data. We also explain and compare two separate methods used to determine SFRs. Results are given in Section~\ref{sect:results} along with a discussion of the possibility of a $L_X$--SFR--Metallicity relation. This is followed by discussion, summary and conclusions in Sections~\ref{sect:discussion} and \ref{sect:conclusions}. All errors are at the 68\% level, unless otherwise stated.


\section{Sample Selection}\label{sect:sample}

\begin{figure}
\centering
\includegraphics[width=0.49\textwidth]{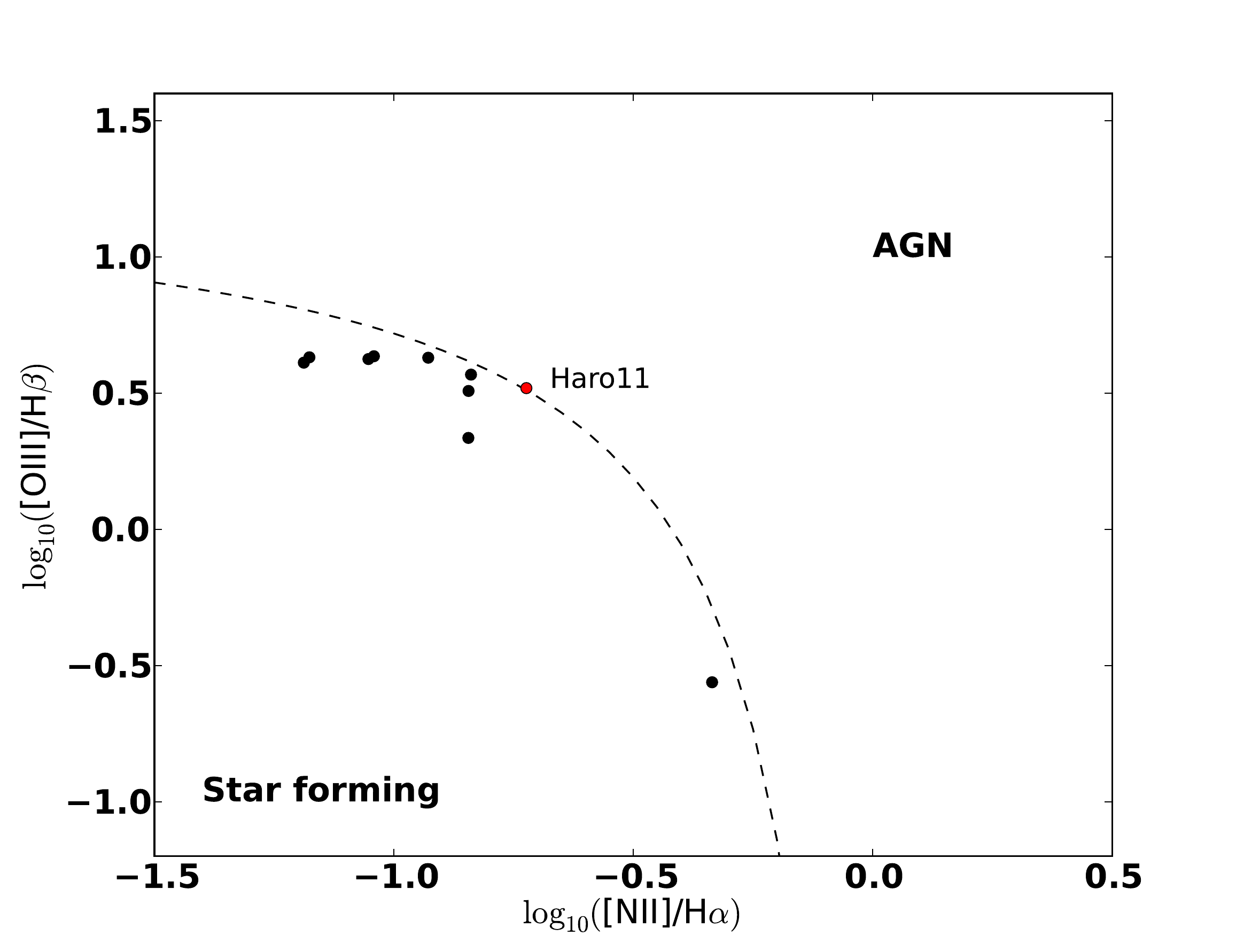}\\
\caption{BPT \protect\citep{baldwin1981} diagram used for AGN vs. starburst diagnostics. The sample of ten LBAs used in this study all fall within the strict starburst region as defined by \protect\cite{kauffmann2003}. Haro~11 did not have SDSS coverage so we used the reported line fluxes of \protect\cite{bergvall2002} to calculate line ratios. All other LBAs have SDSS reported line fluxes.
}\label{fig:agn}
\end{figure}

The sample of LBAs considered here are drawn from the UV imaging survey performed by the \emph{Galaxy Evolution Explorer (GALEX)}. \cite{heckman2005} defined a set of supercompact, UV-luminous galaxies (scUVLGs) at $z<0.3$ that have similar properties to the more distant LBGs. The sample was defined as having FUV luminosities such that $(L_\text{FUV} > 10^{10.3}\, L_{\sun})$, and was later refined by adding the supercompact constraint for surface brightness, $(I_\text{FUV} > 10^9\, L_{\sun} $ kpc$^{-2})$~\citep{hoopes2007}. From this sample of LBAs, five had been observed by Chandra: \cite{grimes2007} studied Haro~11 and \cite{basu-zych2013b} observed four more LBAs for their study. We utilized the \emph{Chandra X-ray Observatory} to observe five more LBAs, bringing our sample size to ten. We required the LBAs to be pure starbursts according to the optical line emission criteria of \cite{kauffmann2003} (BPT-diagram), thereby limiting the likelihood of AGN contamination (see Figure~\ref{fig:agn}). We also used 1.4~GHz spectral luminosities \citep{condon1998,basu-zych2007} to look for clear AGN candidates with $L_{1.4{\rm GHz}} > 10^{24}$~W~Hz$^{-1}$ \citep{alexander2012}. We found values for six out of ten LBAs, all below the AGN cutoff with $L_{1.4{\rm GHz}} < 10^{23}$~W~Hz$^{-1}$. Using SFR estimates from \cite{brinchmann2004} and $L_X$/SFR relation of \cite{mineo2012}, we estimated $L_X$ and selected the four brightest, non-composite LBAs with $12+\log_{10}({\rm O/H}) < 8.4$ and one with near-solar $12+\log_{10}({\rm O/H}) = 8.8$.
Optical spectra and images confirm that the LBAs are similar to LBGs in their SFR, physical size, stellar mass, gas velocity dispersion, and metallicity. High resolution UV imaging shows that UV emission originates in highly compact burst regions in small, clumpy galaxies that are morphologically similar to LBGs~\citep{overzier2009}. Many studies have now established these galaxies as the best known local analogues to LBGs~\citep{basu-zych2007,basu-zych2009,basu-zych2013b,overzier2008,overzier2009,goncalves2010,heckman2011}.

\subsection{Comparison Samples}
Throughout the paper, we use the 29 star-forming spiral and irregular galaxies found in Table~1 of M12 as a comparison sample for the $L_X$--SFR relation. In Section~\ref{sect:plane}, we examine the possible metallicity dependence of the $L_X$--SFR relation. M12 does not directly provide metallicity measurements for their data, so in this case we compare with the results of \cite{douna2015}, who found reported metallicities for 19 of the galaxies in M12. The sample of \cite{douna2015} consists of galaxies from two main sources: \cite{mineo2012} and \cite{brorby2014}. The \cite{douna2015} subset of M12 galaxies (19 out of the original 29), shown in our figures as blue squares, have gas-phase metallicities of $12+\log({\rm O/H})>8$. The remaining set of galaxies in \cite{douna2015} are mainly BCDs from \cite{brorby2014}, and are shown in our figures as green circles (black triangles for those that only have $L_X$ upper limits), and have $12+\log{(\rm O/H)} < 8$.


\begin{table*}
\caption{Sample of Lyman break analogue galaxies. The table includes the RA and DEC (J2000) of each LBA, the redshift (NED), apparent size (radii in arcsec, SDSS isophotal D25), gas-phase metallicity (12+log(O/H)) where solar is 8.69 (BZ13; Pettini~\&~Pagel~(2004)).}
\centering
\begin{tabular}{lllcccccc}
\hline\hline
Name                    & ObsID & Obs. Date   & Exp. Time & RA     & DEC   & $z$   & Apparent size & Metallicity* \\
						&		&				& (ks)		& (J2000)&(J2000) &  & (arcsec) & (12+log(O/H))\\
\hline
Haro~11                 &8175   & 2006-10-28    & 54.0  & 00 36 52.70   & $-$33 33 17.0 & 0.020 & 46.9$\times$46.6              & 8.33$^a$ \\
KUG 0820+282            &13012  & 2011-04-28    & 8.9   & 08 23 54.95   & $+$28 06 21.6 & 0.047 & 13.1$\times$22.4              & 8.23 \\
KUG 0842+527            &16021  & 2013-12-07    & 6.9   & 08 46 02.23   & $+$52 31 59.1 & 0.053 & 29.8$\times$28.9              & 8.80 \\
SHOC 011                &13014  & 2011-10-23    & 19.1  & 00 21 01.03   & $+$00 52 48.1 & 0.098 & 8.2$\times$8.2                & 8.20 \\
SHOC 263                &16018  & 2013-12-22    & 15.5  & 09 38 13.50   & $+$54 28 25.1 & 0.102 & 12.3$\times$8.6               & 8.19  \\
J080619.49+194927.2     &13015  & 2010-12-15    & 19.8  & 08 06 19.50   & $+$19 49 27.3 & 0.070 & 12.5$\times$13.9              & 8.15 \\
J225140.31+132713.3     &13013  & 2011-01-17    & 19.6  & 22 51 40.32   & $+$13 27 13.3 & 0.062 & 10.2$\times$10.9              & 8.15 \\
SHOC 042                &16019  & 2014-01-31    & 22.7  & 00 55 27.46   & $-$00 21 48.6 & 0.167 & 6.9$\times$7.3                & 8.28 \\
J082413.12+433721.0     &16022  & 2013-12-20    & 23.5  & 08 24 13.13   & $+$43 37 21.0 & 0.118 & 8.2$\times$10.4               & 8.35 \\
SHOC 595                &17418  & 2014-09-24    & 13.6  & 23 07 03.75   & $+$01 13 11.2 & 0.126 & 14.7$\times$8.1               & 8.30 \\
\hline
\end{tabular}
\label{tab:sample}
\\
\raggedright
\vspace{1mm}
\textbf{Notes.} Redshifts were taken from NED and were determined from optical emission/absorption lines using SDSS data. For Haro~11, we use the value reported by Bergvall~et~al.~(2000) where they use the LWS on the ISO.\\
* Metallicity is calculated using line ratios from SDSS DR7 following the method outlined by Pettini~\&~Pagel~(2004), using [OIII] $\lambda 5007$, [NII] $\lambda 6584$, H$\alpha$, and H$\beta$ line ratios. All values have 68\% errors of $\pm$0.14.\\
$^a$ Based on reported line fluxes from Bergvall~\&~\"{O}stlin~(2002).
\end{table*}

\section{Observations}\label{sect:observations}
\subsection{X-ray Observations}\label{sect:chandra_obs}
Our sample consists of ten LBAs, all of which have been observed by the \emph{Chandra X-ray Observatory}. Five of these galaxies were specifically observed for this study (ObsID: 16018--16022, 17418). These observations were taken during Observation Cycle 15 between the dates of 2013-12-07 and 2014-09-24. One of the other galaxies was initially observed by \cite{grimes2007}~(Haro~11). \cite{basu-zych2013b} (hereafter, BZ13) observed four more LBAs for their study with observations taking place between 2011-01-17 and 2011-12-15. BZ13 found enhanced X-ray emission with a mean $1.5\sigma$ higher than the \cite{mineo2012} results. 

All observations were obtained using the ACIS-S3 back-illuminated chip aboard Chandra. We reprocessed the level 1 event files using the latest version of CIAO~(4.7.1) and CALDB~(4.6.5). The names, coordinates, redshifts, and metallicities of the galaxies in our sample are listed in Table~\ref{tab:sample}.

\subsection{Infrared and Ultraviolet Observations}
The WISE AllSky Survey \citep{wright2010} provided complete coverage of the sky in four infrared bands. All ten of the galaxies in our sample have unique, coincident WISE sources. We downloaded these data from the archive and converted the $22 \mu$m (WISE band 4) magnitudes to monochromatic fluxes, as outlined by \cite{wright2010}. We used these derived fluxes in our SFR indicators.

A second component of the SFR indicators is the ultraviolet (UV) luminosity.
Images of each galaxy were found in the GALEX archive. We use near UV (NUV, $\lambda_\text{eff} = 2312$\AA ) GALEX data to obtain UV luminosities, as outlined in Section~\ref{sect:SFR}.

\subsection{Metallicity Measurements}\label{sect:metallicity}
The gas-phase metallicity for each galaxy was calculated using line ratios from SDSS DR7 spectral data~\citep{abazajian2009} following the method outlined by \cite{pettini2004}. This method requires emission line measures of [OIII]~$\lambda 5007$, [NII]~$\lambda6584$, H$\alpha$, and H$\beta$. The relation is given by $12+\log_{10}(\text{O/H}) = 8.73-0.32\times$O3N2, where O3N2 $= \log_{10}\{([\text{OIII}]/H\beta)/([\text{NII}]/H\alpha)\}$. This metric is valid for ${\rm O3N2} < 1.9$ with a 68 per cent (95 per cent) confidence interval of $\pm 0.14\ (\pm0.25)$~dex.
Solar gas-phase metallicity is taken to be $12+\log_{10}(\text{O/H}) = 8.69$ \citep{allende-prieto2001,asplund2004}, with an absolute metallicity of $Z_{\sun}=0.02$. One of the galaxies, Haro~11, did not have SDSS coverage. We use the reported line fluxes of \cite{bergvall2002} to calculate gas-phase metallicity in this case. The use of this metallicity measure allows us to directly compare our results with \cite{basu-zych2013b}. 
\begin{figure}
\centering
\includegraphics[width=0.5\textwidth]{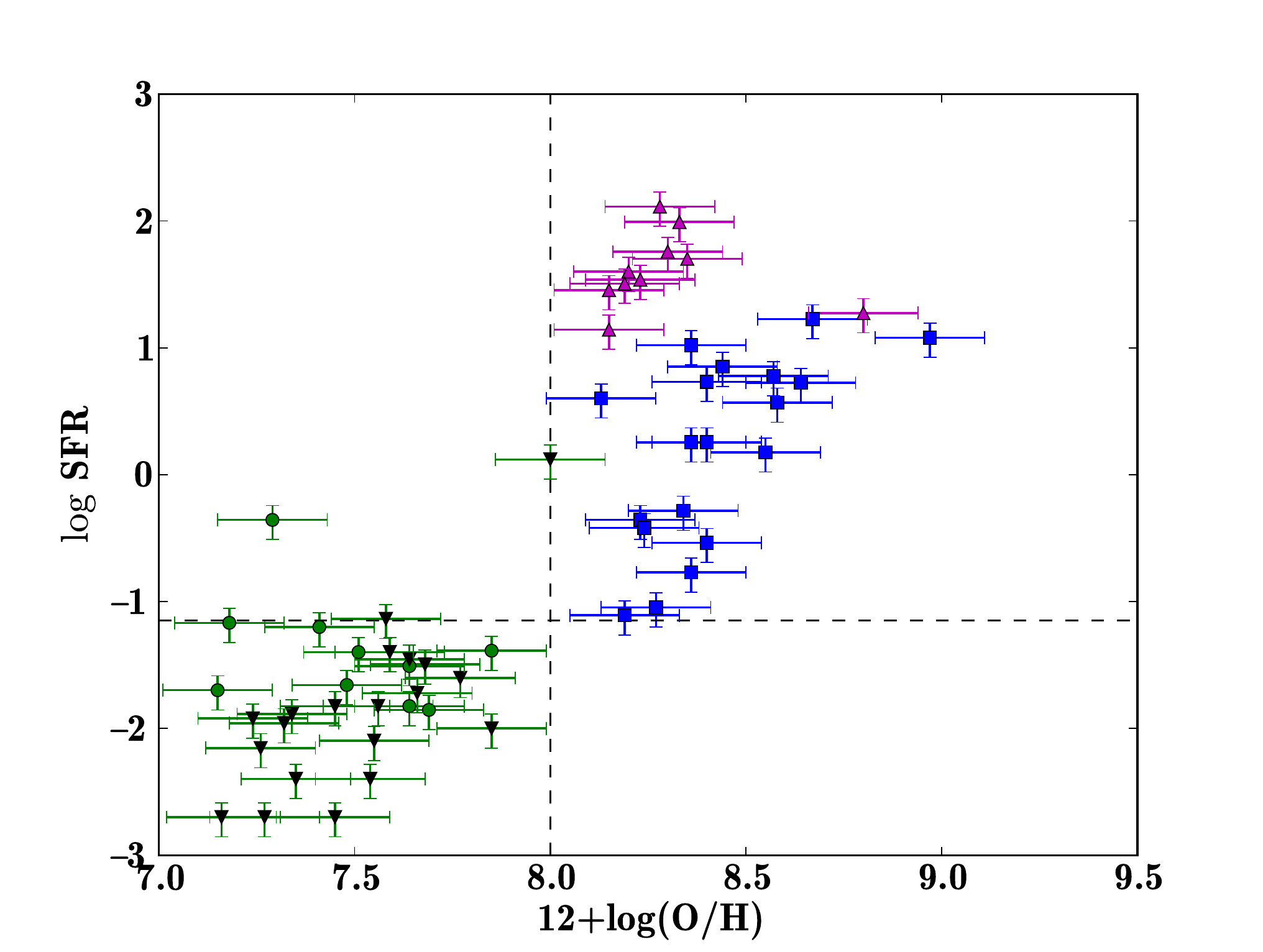}\\
\caption{SFR versus gas-phase metallicity. The comparison samples from \protect\cite{mineo2012} and \protect\cite{brorby2014} in \protect\cite{douna2015} are shown in blue (squares) and green (circles), respectively. The LBA sample of this paper are plotted in magenta (triangles). The galaxies cover distinct regions of the parameter space. We have adjusted the metallicity values of our sample from the O3N2 method of PP04 to the N2 method of PP04 in order to be consistent with the comparison sample from \protect\cite{douna2015}. Error estimates are assumed to be 30\% of SFR value and $\pm0.14$ for $12+\log({\rm O/H})$. The dashed lines separate the SFR--metallicity space into regions within which samples of galaxies have been drawn and those which are relatively sparse. Future observations should sample the empty regions of the parameter space in order to further test the results discussed in this paper.}\label{fig:SFR_metal}
\end{figure}

\section{Procedure}\label{sect:procedure}
The X-ray luminosity for each galaxy was calculated from the net number of counts $(0.5-8$~keV$)$ received within a predefined ellipse, using a background region located on the same CCD chip. The GALEX NUV images corresponding to each X-ray source were used to define these elliptical source regions. The ellipse sizes were defined by varying the size of the source region until the net counts (in the GALEX NUV image) became insensitive to region size. This set the semi-major axis length for the ellipse. The semi-minor axis dimension was determined based on the apparent NUV morphology of the given galaxy. Using these source and background regions, we ran the CIAO tool \texttt{srcflux} to calculate the unabsorbed flux in the $0.5-8$ keV energy band assuming an absorbed power law model. For each source, we assumed a photon index of $\Gamma = 2.0$ and determine $n_H$ by using the \texttt{prop\_colden} tool in CIAO. We chose this method, as opposed to the \cite{basu-zych2013b} method of extracting flux from the 2-10~keV band, because the net counts in the hard band were too few to allow for statistically significant flux values (see Table~\ref{tab:results}). Instead we converted the full energy band flux to the hard band when comparing to BZ13 (conversion factor of 0.645, assuming average $n_H = 3\times 10^{20}~{\rm cm}^{-2}$ and power laws of 1.9 and 2.0 for BZ13 and M12, respectively). Taking band ratios of net counts found in the $0.5-2$ to $2-10$~keV energy ranges, we find our assumption of a power law with $\Gamma = 2.0$ is consistent with all observations.

\subsection{Star Formation Rate Indicator}\label{sect:SFR}
Using UV and IR emission escaping the target galaxy, M12 utilized a SFR indicator for their sample of starburst galaxies based on the work of \cite{bell2003,hirashita2003,iglesias2004,iglesias2006}. The UV emission was determined from \emph{GALEX} NUV background-subtracted intensity maps. The pixel values for the image files report photons~pixel$^{-1}$~s$^{-1}$ corrected for relative response. Within the ellipses described in Section~\ref{sect:procedure}, we converted the net counts to monochromatic flux (at $2312$~\AA ) using the conversion factor\footnote{\scriptsize \url{http://galexgi.gsfc.nasa.gov/docs/galex/FAQ/counts_background.html}} 
\[
C_\text{NUV} = 2.06\times 10^{-16}\ [(\text{erg~cm}^{-2}\text{~s}^{-1}\text{\AA}^{-1})/(\text{ph s}^{-1})].
\]
M12 used \emph{Spitzer} data to determine the IR component of SFR. However, for our sample of LBAs only four galaxies have \emph{Spitzer} observations. Instead we used the wider coverage of \emph{WISE} to determine the IR luminosity of each galaxy in our sample. Using the procedure outlined by \cite{wright2010}, we converted WISE band 4 (22-$\mu$m) magnitudes to monochromatic luminosities. Following BZ13, we estimated $L_\text{IR}$ in the 8--1000~$\mu$m band using the WISE monochromatic luminosities and the SED templates of \cite{chary2001}.

The total SFR used by M12 is given by 
\begin{equation}
\text{SFR}_\text{tot} = \text{SFR}^0_\text{NUV} + (1-\eta) \text{SFR}_\text{IR},
\end{equation}
where SFR$^0_\text{NUV}$ and SFR$_\text{IR}$ are obtained assuming a 100~Myr old stellar population with constant SFR and a Salpeter initial mass function (IMF) from $0.1-100$~M$_{\sun}$:
\begin{align*}
\text{SFR}^0_\text{NUV} &= 1.2\times 10^{-43} L_\text{NUV,obs} (\text{erg s}^{-1}),\\
\text{SFR}_\text{IR} &= 4.6\times 10^{-44} L_\text{IR} (\text{erg s}^{-1}),
\end{align*}
where $L_\text{NUV,obs}$ is the observed luminosity (at 2312~\AA ), uncorrected for dust attenuation, and $L_\text{IR}$ is the total IR luminosity (8--1000~micron). The correction factor, $\eta$, accounts for the fraction of IR emission due to old stars versus new stars in the galaxy. For starburst galaxies, almost all of the IR emission is due to recent star formation and so we assume $\eta \approx 0$~\citep{hirashita2003}.

\cite{basu-zych2013b} used a UV+IR determined SFR described by \cite{bell2005} assuming a 100~Myr old stellar population with constant SFR and a Kroupa IMF from $0.1-100$~M$_{\sun}$.
Comparing the SFR coefficients from BZ13 and the ones used here, we find a conversion factor of ${\rm SFR_{M12}/SFR_{BZ13}} = 1.79$. This is consistent with our observed relation of ${\rm SFR_{M12}/SFR_{BZ13}} = 1.783\pm 0.009$ from the LBA data.
\begin{table*}
\caption{Sample of Lyman break analogue galaxies. Stellar masses were taken from the MPA-JHU galaxy catalog for SDSS DR7$^a$ with a median uncertainty of 0.09~dex (68\%), star formation rate in $M_{\sun}$/yr, X-ray flux (and luminosity) in the 0.5-8 keV energy range in units of $10^{-15}$~erg~cm$^{-2}$~s$^{-1}$ assuming a power law with $\Gamma = 2.0$, and the exposure time in kiloseconds. Net count errors are combined Poisson errors at the 68\% level.}
\centering
\begin{tabular}{lcccccccc}
\hline\hline
Name                    &$\log M_\star$& SFR$_{\rm BZ13}$   & SFR$_{\rm M12}$   & Net Counts  & Net Counts   &  $N_H$  & Flux  & $L_X$   \\
                        &$(M_{\sun})$& $(M_{\sun}$ yr$^{-1})$ &$(M_{\sun}$ yr$^{-1})$    & $(0.5-8\,$keV$)$ &$(2-10\,$keV$)$ & ($10^{20}$cm$^{-2}$)& ($10^{-15}$ erg cm$^{-2}$ s$^{-1}$)    & ($10^{41}$ erg s$^{-1}$)   \\
\hline
Haro~11$^\dagger$       & 9.84 & 54.6  & 98.1  & 1904$\pm$57  & 466$\pm$50	& 1.88  &    245.0$\pm7.3$ & 2.22$\pm0.07$    \\
KUG 0820                & 8.61 & 19.3  & 34.4  & 56.1$\pm$8.6 & 9.3$\pm$6.4	& 4.85  &    52.0$\pm8.0$  & 2.7$\pm0.4$    \\
KUG 0842                & 11.08& 10.6  & 18.8  & 30.9$\pm$8.5 &   $<$8.8		& 3.07  &    39.6$\pm10.9$ & 2.6$\pm0.7$    \\
SHOC 011                & 9.29 & 22.5  & 39.9  & 32.4$\pm$6.3 & 5.7$\pm$4.4	& 2.62  &    13.5$\pm2.6$  & 3.3$\pm0.6$    \\
SHOC 263                & 9.37 & 18.0  & 32.0  & 16.8$\pm$4.9 & 6.4$\pm$4.4	& 1.61  &    9.4$\pm2.7$   & 2.6$\pm0.7$    \\
J080619.49              & 9.26 & 16.0  & 28.5  & 27.1$\pm$7.2 &   $<$13.3	& 3.44  &    10.8$\pm2.9$  & 1.3$\pm0.3$    \\
J225140.31              & 9.15 & 7.85  & 13.9  & 52.1$\pm$8.3 & 12.5$\pm$6.3	& 4.85  &    21.5$\pm3.4$  & 2.0$\pm0.3$    \\
SHOC 042                & 9.66 & 72.3  & 129.9 & 13.6$\pm$4.5 &   $<$3.0		& 2.89  &    5.4$\pm1.8$   & 4.3$\pm1.4$    \\
J082413.12              & 10.24& 28.2  & 50.3  & 39.9$\pm$7.2 &   $<$5.1	& 4.21  &    15.4$\pm2.8$      & 5.6$\pm1.0$    \\
SHOC 595                & 9.48 & 32.0  & 57.3  & 17.6$\pm$5.1 & 8.7$\pm$4.9	& 4.73  &    12.4$\pm3.6$  & 5.2$\pm1.5$    \\
\hline
\end{tabular}
\label{tab:results}
\\
\raggedright
$^\dagger$ Source region ellipse for Haro 11 encompasses all three 'knots' of bright emission.\\
$^a$ \url{http://wwwmpa.mpa-garching.mpg.de/SDSS/DR7/}
\end{table*}

\section{Results}\label{sect:results}
\begin{figure*}
\centering
\includegraphics[width=0.8\textwidth, natwidth=610, natheight=642]{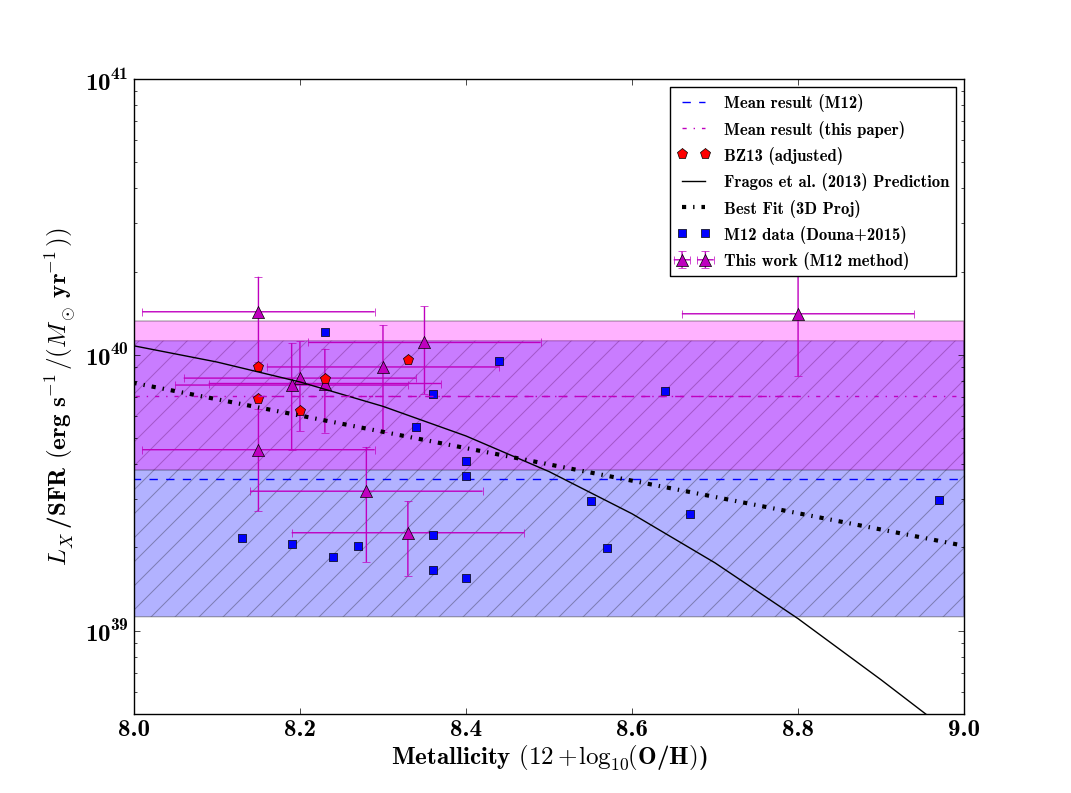}\\
\caption{Plot showing the $L_X$/SFR relation as discussed in the Results section. The red pentagons are the reported $L_X$/SFR values from BZ13 adjusted to compare to M12. The dashed line (blue) and the blue band (hatched) represent the \protect\cite{mineo2012,mineo2012b} results for unresolved sources in the $0.5-8$~keV range and M12-equivalent SFRs: $L_X$/SFR $= 3.9\times 10^{39}$ (erg s$^{-1}$)/($M_{\sun}$ yr$^{-1}$) with dispersion of 0.34~dex. The magenta dot-dashed line and magenta band are the mean and dispersion we calculate using the M12 SFR indicator ($L_X$/SFR $= 7.08\times 10^{39}$, $\sigma = 0.27$~dex) for our data (magenta triangles). The black solid line is the prediction from \protect\cite{fragos2013b} and the black dot-dashed line represents the projected fit from Section~\ref{sect:plane}.
\textbf{Note:} There is a significant disagreement in $L_X$/SFR between the BZ13 published value and our reported value for the galaxy with metallicity of 8.33 (Haro~11). This disagreement largely arises from SFR value. BZ13 reports a value of $10.88\ M_{\sun}$ yr$^{-1} $, whereas we find a value of $54.6\ M_{\sun}$ yr$^{-1}$. The most likely reason for this is that we use raw WISE (IR) and GALEX (UV) data, and BZ13 uses the reported value of $L_{\rm IR}$ from \protect\cite{grimes2007}, who used IRAS multiband data. }\label{fig:Lx_SFR}
\end{figure*}

\noindent From our derived results for X-ray luminosity and SFRs, which are summarized in Table~\ref{tab:results}, we determined the $L_X$--SFR relation using a logarithmic least squares fitting technique, assuming a linear relation (i.e., we apply least-squares fitting to the logarithmic values of $L_X$/SFR). 
We find the relation
\begin{align*}
L_{0.5-8\text{keV}}(\text{erg s}^{-1}) &= 7.10\times 10^{39} \text{ SFR}\ (\text{M}_{\sun} \text{ yr}^{-1}),\\
\log_{10} L_{0.5-8\text{keV}} &= 39.85(\pm 0.10) + \log_{10} \text{SFR}\ ,
\end{align*}
with a dispersion of $\sigma = 0.25$~dex (see Figure~\ref{fig:Lx_SFR}). In this case, the X-ray luminosity was calculated from the net counts in the $0.5-8$~keV range using $\Gamma = 2.0$ plus the Galactic absorption along the line of sight (see Table~\ref{tab:results}).

We wish to compare these results to the results of M12, but their $L_X-$SFR relation uses only resolved XRB point sources, whereas our results use the total, unresolved X-ray emission from the galaxies. In a follow-up paper, \cite{mineo2012b} determined the X-ray contribution from the unresolved, diffuse component of their sample of galaxies. They found that this also correlates linearly with SFR such that,
\[
L^{\rm diff}_{0.5-2~{\rm keV}} ({\rm erg s}^{-1}) = 8.3\times 10^{38}\ {\rm SFR} ({\rm M_{\sun}~yr}^{-1}),
\]
with a dispersion of $\sigma = 0.34$~dex. $L^{\rm diff}_{0.5-2~{\rm keV}}$ is the unresolved emission, from which the contribution of backgrounds and unresolved HMXBs was subtracted. The sample consisted of a subset of 21 out of the original 29 galaxies used in the M12 Paper I. Expanding this relation to the $0.5-8$~keV range, using the two-component thermal plus power law model discussed in \cite{mineo2012b}, we find a conversion factor of 1.17. We can add this to the resolved XRB $L_X-$SFR relation of M12 to get,
\begin{align*}
L^{\rm XRB+diff}_{0.5-8~{\rm keV}} ({\rm erg s}^{-1}) &= 3.9\times 10^{39}\ {\rm SFR} ({\rm M_{\sun}~yr}^{-1}),\\
\log_{10} L_{\rm X} &= 39.59 + \log_{10} \text{SFR}\ ,
\end{align*}
with dispersion $\sigma = 0.34$~dex (Figure~\ref{fig:Lx_SFR}). Thus, our result shows a factor of $1.8^{+0.8}_{-0.6}$ increase in X-ray luminosity with respect to SFR.


We have seen that the sample of LBA galaxies exhibit an elevated $L_X$/SFR relation as compared to the results for spiral and irregular starburst galaxies \citep{mineo2012}. Comparing these two samples we calculate the significance of this elevated X-ray luminosity using the two-sample t-test method (Welch statistic), assuming the null hypothesis (no difference in mean values). We compare our results $(\mu=39.85, \sigma=0.25, n=10)$ with M12 $(\mu=39.59,\sigma=0.34,n=29)$ and get a probability of $0.018~(2.4\sigma)$ that the two samples are drawn from the same population. 
Comparing the difference between the observed values of $L_X$/SFR for the 10 LBAs and the 29 galaxies from M12, we find a one-tailed significance of 98.2\% using a permutation test~\citep{fisher1935,dwass1957,barnard1963}, consistent with the t-test result. Therefore, the LBA population does show significant enhancement of $L_X$/SFR compared to that for M12.

If we restrict ourselves to the metallicity range $8.0 < 12+\log({\rm O/H}) < 8.5$, the common ground for the LBAs and M12 sample, we find a reduced significance of $96.2$\% from the two-sample t-test. This may suggest an $L_X$/SFR enhancement in the LBAs not attributable to metallicity effects.

\subsection{Possible $L_X$--SFR--Metallicity Plane for Star Forming Galaxies}\label{sect:plane}
In this section we discuss a possible plane describing the X-ray luminosity of a star-forming galaxy dominated by HMXB $({\rm SFR}/M_\star > 1\times 10^{-10}~{\rm yr^{-1}})$. M12 did not publish associated metallicities for their sample. However, \cite{douna2015} found metallicities for 19 out of 29 galaxies using SDSS optical data consistent with the measurement technique used in this paper. \cite{douna2015} compared the $L_X$/SFR of BCDs found in \cite{brorby2014} and star-forming galaxies in M12 versus metallicity and found an enhancement of a factor of ten. We use the sample of \cite{douna2015} as our comparison sample for metallicity dependence. We split the sample into the $12+\log({\rm O/H})>8$ M12 subset (blue squares) and the remaining galaxies ($12+\log({\rm O/H})<8$), which consist mostly of BCDs found in \cite{brorby2014} (green circles, black triangles ($L_X$ upper limits only)). The errors on luminosity $(L_X)$ for the comparison sample are estimated from threshold luminosity and measured luminosity (Table~1 in \cite{douna2015}), assuming Poisson errors. All data from \cite{douna2015} have been corrected to be consistent with the measurement techniques used in this paper, as described in the previous section. We exclude data that have only upper limits on $L_X$ from our fitting procedure.

Testing for a correlation between metallicity and $L_X$/SFR over all 39 galaxies in the combined data set, we find a negative correlation at the 99.9\% significance level, based on a Spearman's rank correlation test. 

\begin{figure*}
\centering
\includegraphics[width=0.49\textwidth]{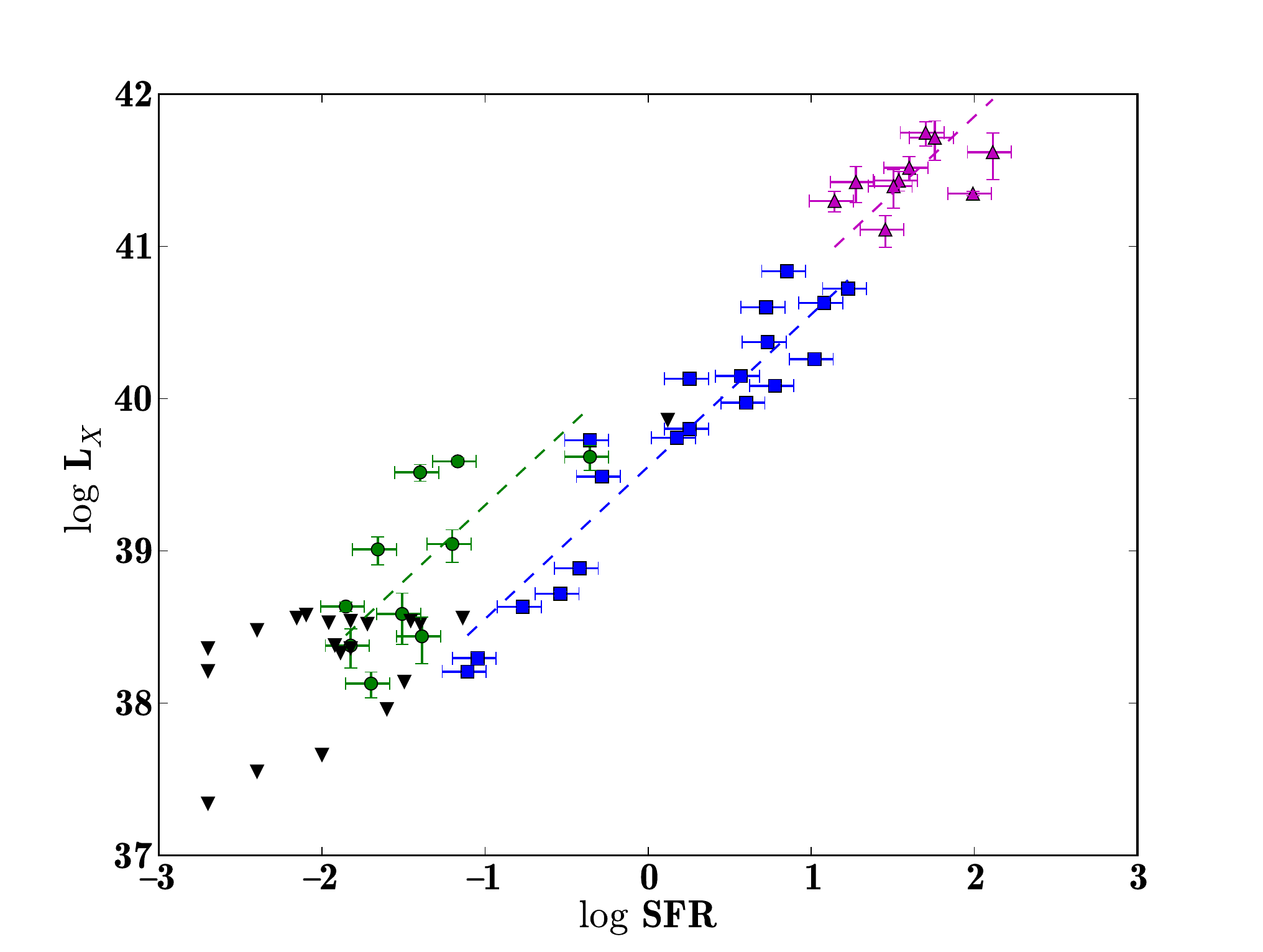}\includegraphics[width=0.50\textwidth, natwidth=610, natheight=642]{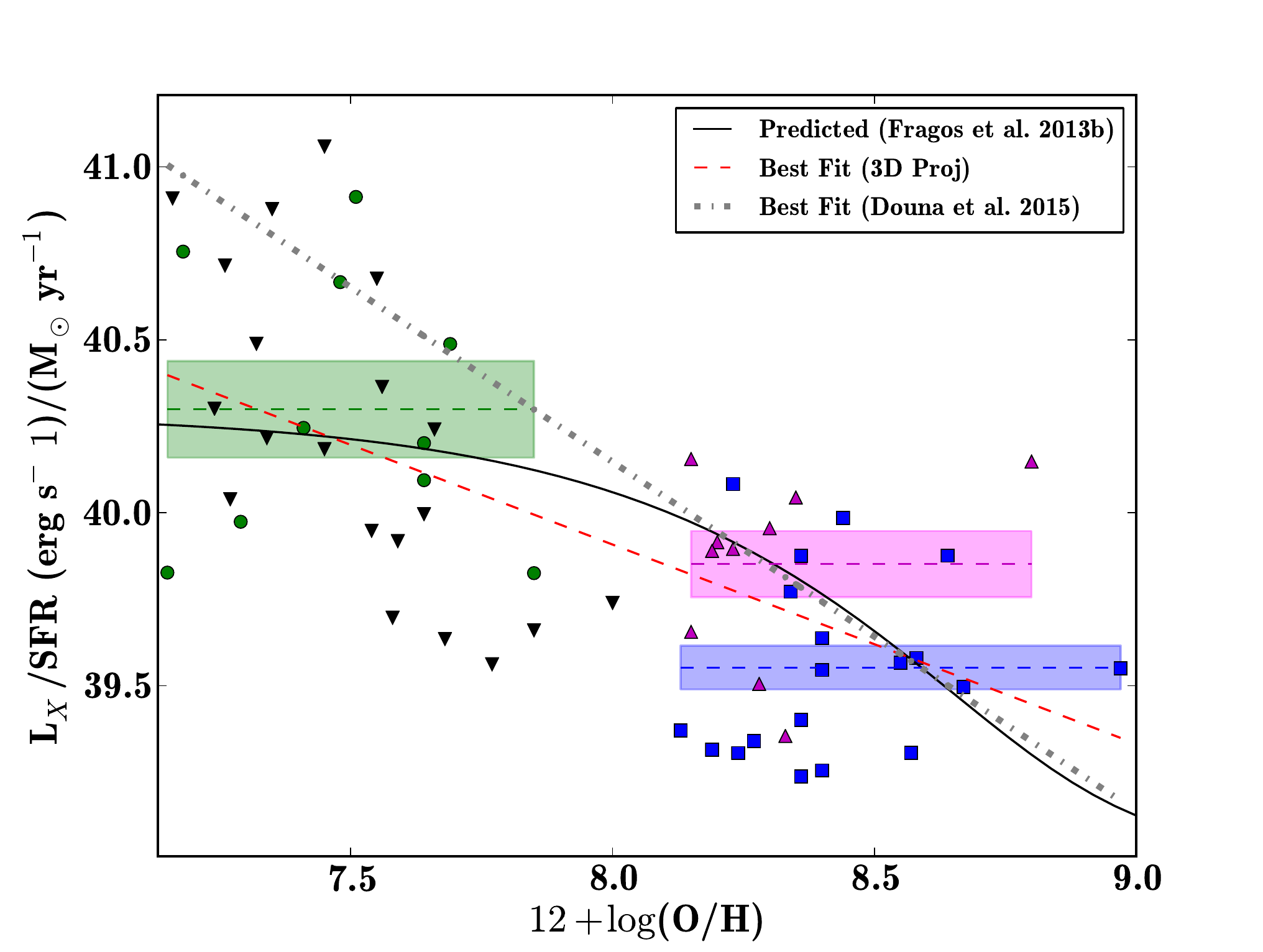}\\
\includegraphics[width=0.6\textwidth, natwidth=610, natheight=642]{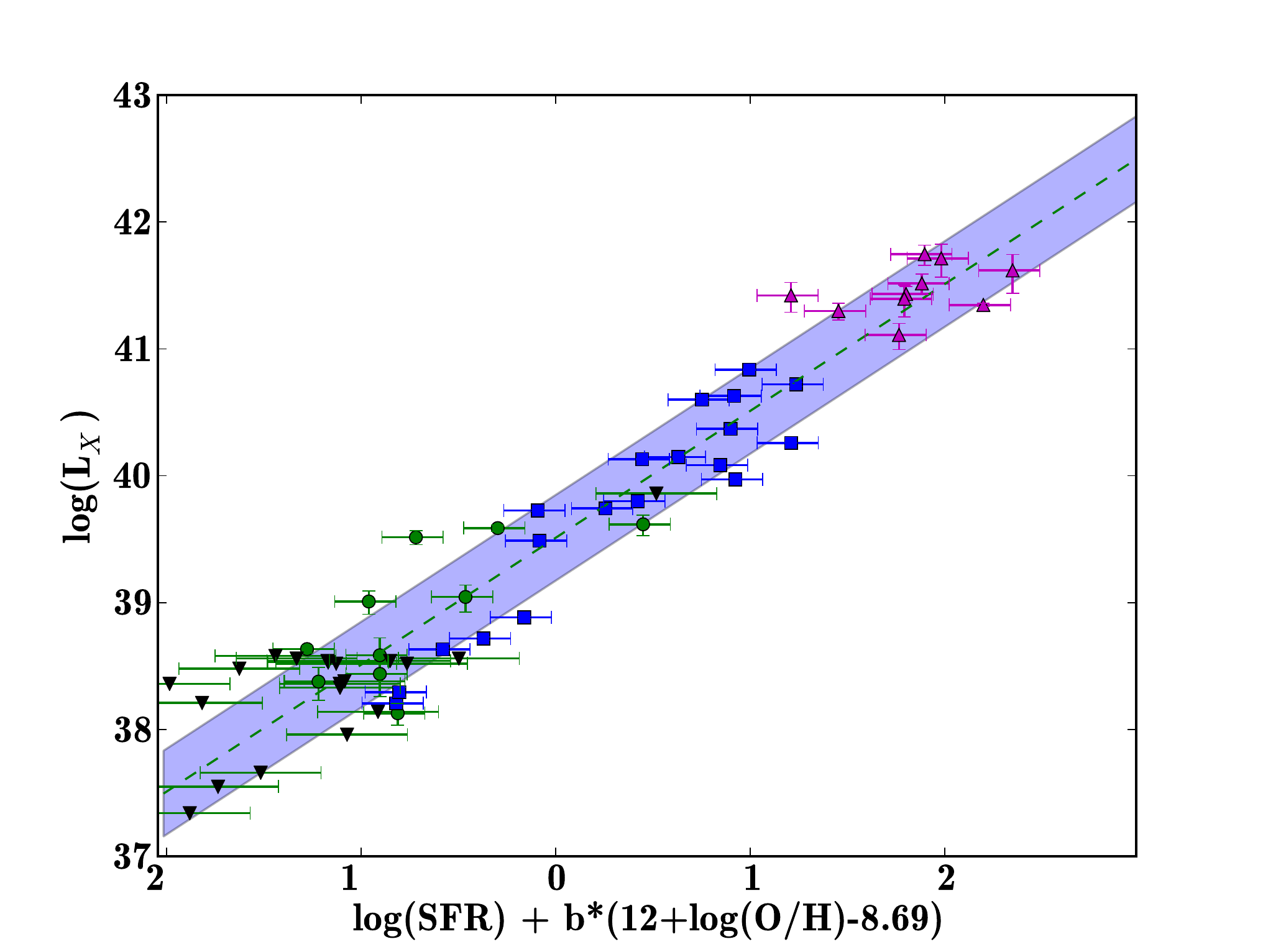}
\caption{Various projections of $L_X$--SFR--Metallicity. The sample of \protect\cite{douna2015} consisting of mostly BCDs are shown as circles (green) with upper limits given by upside-down triangles (black). The subset of \protect\cite{douna2015} that includes \protect\cite{mineo2012} spiral and irregular galaxies are plotted as squares (blue). The set of LBAs from this paper are plotted as triangles (magenta). \emph{Upper Left}: $L_X$--SFR relation for the various subsets of star-forming galaxies. The dashed lines represent linear fits to the $L_X$--SFR relations. We find $\log(L_X/{\rm SFR}) = 40.30\pm0.14, 39.59\pm0.06, 39.85\pm0.10$ for the green, blue, and magenta data points, respectively. The BCDs and LBAs exhibit enhanced X-ray emission relative to the spirals and irregulars. \emph{Upper Right}: The $L_X$/SFR--Metallicity relation shows a negative correlation with metallicity with slope $= -0.59\pm0.13$ (dashed, red line). The linear fits from the $L_X$--SFR relations are plotted as dashed lines with surround shaded regions representing the errors on the means. The black solid line is the prediction from \protect\cite{fragos2013b} and the grey dot-dashed line represents the projected fit from \protect\cite{douna2015} (see Section~\ref{sect:plane}). \emph{Lower}: The best fitting $L_X$--SFR--Metallicity relation is shown (dashed line) where $a=1$ is assmumed and we find $b=-0.59\pm 0.13$, $c=39.49\pm 0.09$, $\chi^2/{\rm d.o.f.} = 91.39/37$, and dispersion $\sigma = 0.34$~dex. Allowing $a$ to vary, we find $a=1.03\pm 0.06$, $b=-0.64\pm 0.17$, and $c=39.46\pm 0.11$ with $\chi^2/{\rm d.o.f.} = 88.2/36$ and dispersion $\sigma = 0.34$~dex. The forced fit ($a=1$) is consistent with and indistinguishable from the model with $a$ as a free parameter. Therefore, the model with $a$ set to unity is favored and is conistent with previous studies that find a linear relation between $L_X$ and SFR. }\label{fig:plane_results}
\end{figure*}

We then fit a $L_X$--SFR--Metallicity relation of the form
\[
\log\left(\frac{L_X}{\rm erg~s^{-1}}\right) = a \log\left(\frac{\rm SFR}{\rm M_{\sun}~yr^{-1}}\right) + b \log\left(\frac{\rm (O/H)}{\rm (O/H)_{\sun}}\right) + c,
\]
where $L_X$ is measured in the $0.5-8$~keV range and $a=1.03\pm0.06$, $b=-0.64\pm0.17$, and $c=39.46\pm0.11$ with $\chi^2/{\rm d.o.f.} = 88.2/36$. The dispersion about this best fit is $0.34$~dex. Since $a$ is consistent with being linear, we force the $L_X$--SFR relation to be linear by setting $a=1$. Doing this we find $b=-0.59\pm0.13$, $c=39.49\pm0.09$, and $\chi^2/{\rm d.o.f.} = 91.39/37$ with a dispersion of $0.34$~dex. This model provides an equally good fit as the model with $a$ as a free parameter. The lower plot in Figure~\ref{fig:plane_results} shows this result. Thus, we confirm that $L_X$ is linearly, positively correlated with SFR and we find that $L_X$ is negatively correlated with metallicity. \cite{douna2015} find a fit, with $a$ set to unity, where $b=-1.01$ and $c=39.26$. Correcting for unresolved vs. resolved emission, their fit corresponds to $c=39.45$ which is consistent with our result. The upper right plot in Figure~\ref{fig:plane_results} shows a projection of the $L_X$--SFR--Metallicity plane for which SFR is forced to be linear. The red, dashed line is a projection of our fitting result and the gray, dot-dashed line is that of \cite{douna2015}. The solid black line shows the prediction of \cite{fragos2013b}, which provides a fit with $\chi^2/{\rm d.o.f.} = 115.4/38$ with a dispersion of $0.38$~dex.

\section{Discussion}\label{sect:discussion}

\cite{linden2010} used population synthesis to show that for star forming regions with $Z \lesssim 0.2\ Z_{\sun}$, the bright HMXB population would be enhanced. \cite{fragos2013b} also used population synthesis and simulations to show that the X-ray luminosity per SFR increases by nearly a factor of ten for metallicities less than 10 per cent solar. \cite{fragos2013b} fit a polynomial to the X-ray luminosity per SFR as a function of metallicity. Using their parameters and solar values of $12+\log_{10}({\rm O/H})=8.69$ and $Z_{\sun} = 0.02$, we calculated an expected mean value of $\log_{10} (L_X$/SFR$) = 39.86$ in the $2-10$~keV range. This is completely consistent with our results. 

The possible existence of a $L_X$--SFR--Metallicity plane is intriguing but not unexpected from simulations \citep{linden2010,fragos2013b}. From Figure~\ref{fig:SFR_metal}, one can see that the three distinct populations of galaxy types (BCDs, spirals, LBAs) occupy three distinct regions of the SFR--metallicity space. Thus the existence of a plane may arise from different $L_X$--SFR relations for these galaxy types and not necessarily the metallicity. 
The fact that LBAs have marginally enhanced X-ray emission relative to the M12 sample in the same metallicity range may suggest that other galactic properties, such as HI gas fraction, affect the $L_X$--SFR relation.
A study of BCDs or LBAs over a wider range in metallicity and SFR (unsampled regions of Figure~\ref{fig:SFR_metal}) could shed light on whether the $L_X$--SFR--Metallicity relation holds or if other properties need to be considered. Expanding the metallicity range of each galaxy type would allow for a better test of the correlations seen between X-ray luminosity, SFR, and metallicity as it provides a more uniform sample.

We find that the \cite{fragos2013b} prediction (black line, Figure~\ref{fig:plane_results}) provides a fit that is not significantly worse than the power law relation (red dashed line). From Figure~\ref{fig:plane_results} one can see that this is not surprising since the Fragos prediction straddles the straight power law fit. Neither the power law model nor the Fragos model are physically motivated but instead are parameterized fits.


\section{Summary and Conclusions}\label{sect:conclusions}
Previous studies of local analogues to early Universe galaxies have shown enhanced X-ray binary populations~\citep{mapelli2010, kaaret2011, prestwich2013, brorby2014, douna2015} and enhanced total X-ray luminosities~\citep{basu-zych2013b} relative to the SFR of the respective galaxies. We continued this line of study by examining the X-ray luminosity per SFR of a sample of ten local analogues to the high redshift Lyman Break Galaxies. Five of the LBAs we studied had been previously observed and analysed by \cite{basu-zych2013b}. The other five LBAs were observed by Chandra specifically for this study. Our results may be summarized as follows:
\begin{enumerate}
  \item The $L_X$/SFR relation for our sample is given by $L^\text{XRB}_{0.5-8\text{keV}}(\text{erg s}^{-1}) = 7.08\times 10^{39} \text{ SFR}\ (\text{M}_{\sun} \text{ yr}^{-1}),$ with a dispersion of $\sigma = 0.25$~dex. Compared to the M12 result for unresolved sources $(3.9\times 10^{39})$, we find a factor of $1.8^{+0.8}_{-0.6}$ increase in $L_X$/SFR with a chance probability of 0.018 that these two samples were drawn from the same population (a significance of $2.4\sigma$). 
  \item We find that when including gas-phase metallicity, the above result is consistent with the prediction of \cite{fragos2013b} based on populations synthesis models (Figure~\ref{fig:Lx_SFR}). For an average metallicity of $12+\log_{10}({\rm O/H}) = 8.3$ or $Z \approx 0.4\, Z_{\sun}$, \cite{fragos2013b} predicts a value for $\log_{10} (L_{2-10 {\rm keV}}^X$/SFR$) = 39.86$, assuming a solar oxygen abundance of $8.69$. This is consistent with our observed value of $39.85\pm0.10$.
  \item Fitting the $L_X$--SFR--Metallicity relation with the form $\log\left(L_X\right) = \log\left({\rm SFR}\right) + b \log\left({\rm (O/H)}/{\rm (O/H)_{\sun}}\right) + c$ gives $b=-0.59\pm 0.13$, $c=39.49\pm 0.09$, and dispersion $\sigma = 0.34$~dex.
\end{enumerate}

These three results provide further evidence that metallicity plays a major role in the evolution of the $L_X$/SFR relation for galaxies. Using local galaxies as analogues suggests that the large, proto-galaxies and the small dwarf galaxies of the early Universe had enhanced X-ray emission due mainly to their low metallicities. The effects of enhanced X-ray emission before and during the Epoch of Reionization would be increased heating of the intergalactic medium, thereby resulting in an earlier onset in the rise of the 21~cm power during this epoch. Many simulations have been done to predict these effects on the redshifted 21~cm line \citep[e.g.,][]{mcquinn2012,mesinger2013}. Using the results of \cite{mirabel2011} (see their Figure~2), a factor of ten increase in the $L_X$/SFR relation (as suggested by the results of \cite{brorby2014,douna2015} and consistent with the predictions of \cite{fragos2013b}) would result in a weakening of the redshifted 21~cm brightness temperature by a factor of 2-3 (or $\Delta(\delta{\rm T_b}) \approx 60 {\rm mK}$). This also results in an earlier onset of the rise of X-ray heating with $\Delta z \approx 2$. Understanding the effects of metallicity on HMXB formation is important for understanding future 21~cm observations of the Epoch of Reionization. 
\section*{ACKNOWLEDGEMENTS}
We thank the anonymous referee for helpful comments and suggestions that greatly improved the manuscript.
The scientific results reported in this article are based on observations made by the Chandra X-ray Observatory. Support for this work was provided by the National Aeronautics and Space Administration through Chandra Award Number G04-15085X issued by the Chandra X-ray Observatory Center, which is operated by the Smithsonian Astrophysical Observatory for and on behalf of the National Aeronautics Space Administration under contract NAS8-03060.


\end{document}